\documentclass[a4,draftcls,12pt,onecolumn]{IEEEtran}
\usepackage{cite}
\usepackage{epsfig,graphics,mathrsfs,amssymb,amsmath,bm,color,yfonts,txfonts}
\usepackage{subfigure}
\usepackage{algorithm2e}

\begin{document}

\title{Convex Approximated Weighted Sum-Rate Maximization for Multicell Multiuser OFDM }
\author{Mirza Golam Kibria and Hidekazu Murata\\
%\thanks{This present study is supported by the Strategic Information and Communications R\&D Promotion Programme (SCOPE) of the Ministry of Internal Affairs and Communications, Japan.}
[2mm]
Graduate School of Informatics, Kyoto University, Kyoto, Japan\\
contact-h25e@hanase.kuee.kyoto.u-ac.jp}

\maketitle

\begin{abstract}
This letter considers the weighted sum-rate maximization (WSRMax) problem in downlink multicell multiuser orthogonal frequency-division multiplexing system. The WSRMax problem under per base station transmit power constraint is known to be NP-hard, and the optimal solution is computationally very expensive. We propose two less-complex suboptimal convex approximated solutions which are based on sequential parametric convex approximation approach. We derive provably faster convergent iterative convex approximation techniques that locally optimize the weighted sum-rate function. Both the iterative solutions are found to converge to the local optimal solution within a few iterations compared to other well-known techniques. The numerical results demonstrate the effectiveness and superiority of the proposed approaches.
\end{abstract}
\IEEEpeerreviewmaketitle
%========================================================
\section{Introduction}
\vspace{-2mm}
\label{section:chapter4-1}
%===================================
Weighted sum-rate maximization (WSRMax) is a fundamental element in many cellular network design problems. Unfortunately, the general WSRMax problem is NP-hard \cite{Luo}, therefore, very difficult to solve. Globally optimal solutions are computationally very inefficient and inapplicable in practical. Therefore, it is intuitive to pursue efficient algorithms for WSRMax, which are even though suboptimal, perform competently in practice. Beamforming designs based on necessary optimality conditions have been thoroughly studied in \cite{Venturino,Wang}. Interestingly, in \cite{Joshi}, it has been shown that the performances of the suboptimal techniques achieving necessary optimality conditions, are indeed very close to the global optimal designs.

%An iterative coordinated beamforming design based on Karush-Kuhn-Tucker (KKT) optimality conditions is proposed in \cite{Venturino}, which is not provably convergent.
In  \cite{Joshi}, an optimal solution for the WSRMax problem based on brach-and-bound (BB) method is proposed. In \cite{Wang}, the authors proposed an alternating maximization (AM) algorithm that is reliant on alternate updating of the beamforming vectors and a closed-form posterior conditional probability. However, the solution is not provably convergent. A weighted minimum mean-square error (WMMSE) based solution is proposed in \cite{Christensen}, which exploits the relationship between the WMMSE and weighted sum-rate. Unfavorably, both of these iterative WSRMax optimization approaches show relatively slower convergence. Recently, a sequential parametric convex approximation (SPCA) based second-order cone program (SOCP) formulation for WSRMax problem is proposed in \cite{Tran}. An efficient multi-carrier extension of \cite{Tran} is proposed in \cite{Mirza}, which further solves the two important numerical instability issues related to algorithm implementation of \cite{Tran}. For a large system (with large number of cells and subcarriers), the solution in \cite{Mirza} is still not effective in terms of problem formulation because of equivalent SOCP constraints transformation of a large number of optimization variables and nonlinear constraints. A similar convex approximation approach known as soft interference nulling (SIN) is proposed in \cite{Chris}, which is based on convex relaxation of the rate function. However, both \cite{Mirza} and \cite{Chris}, arrive at more complex problem formulations.

In this letter, we propose two convex approximated iterative solutions for the WSRMax problem in multicellular multiuser orthogonal frequency division multiplexing (MU-OFDM) system. Like \cite{Tran} and \cite{Mirza}, our iterative designs also exploit the SPCA technique \cite{Beck} that leverages convex approximation of the nonconvex WSRMax problem. However, unlike \cite{Mirza}, our proposed solutions employ different approaches to convexify the problem with less-complex problem formulation that do not need to perform equivalent SOCP transformations for a large number of optimization variables, and the use of differential operator to perform linear approximation is not required. This SPCA based convex approximated WSRMax algorithm yields local optimal solution by iteratively solving the approximated convex problem. Numerical results demonstrates the competency of the proposed WSRMax solutions in terms of convergence rate, computational complexity and problem formulation over other WSRMax solutions.
%\footnote{Note that the non-overlapping subcarrier scheduling in each cell does not restrict the applicability of the proposed algorithm for multiple active users in one subcarrier in one cell.}
%\textit{Notations:} $(\cdot)^{\rm{H}}$ and $(\cdot)^{\rm{T}}$ denote Hermitian transpose and transpose operations, respectively; Gaussian distribution of real/complex random variable with mean $\mu$ and variance $\sigma^2$ is defined as $\mathcal{RN}$/$\mathcal{CN}(\mu,\sigma^2)$; 
%Bold lower/upper case letter defines a vector/ matrix; $\mathbb{C}$ defines a complex space.
%$|\cdot|$ refers to the absolute value. For example, let $t=69$, $M=3$ and $N=64$. Under the current structure of the set $L$, we get $m=2$, $n=5$. Therefore, $L_{69}$ denotes the index set $\{k, m, n\}=\{k, 2, 5\}$. 
\vspace{-1mm}
%===================================
\section{Signal Model and Problem Statement}
\vspace{-2mm}
\label{section:chapter4-2}
%\subsection{System Model}
%===================================
Consider an interference-limited cellular system with $M$ cells and $K$ single-antenna users per cell. All coordinated base stations (BSs) are equipped with $N_\mathrm{Tx}$ antennas. OFDMA scheme with $N$ subcarriers over a fixed frequency band is employed, while the subcarrier assignments among users within each cell are non-overlapping. Therefore, the users do not experience any intra-cell interference. $f(m,n)$ is the assignment function that determines the downlink user scheduling. The assignment of user $k$ from cell $m$ on subcarrier $n$ is defined by $k=f(m, n)$. Let $\mathcal{M}\triangleq\{1,2,\cdots, M\}$. The received signal at user $k$ is given by
\begin{equation}
\label{optz1}
{{y}_{k,m,n}}={{\bm{h}}_{k,m,n}^{\rm{H}}}{{\bm{w}}_{k,m,n}}{{d}_{k,m,n}}+\hspace{-4mm}\sum\limits_{\begin{smallmatrix}
 {m}'\in \mathcal{M}\backslash m \\
 {k}'=f(m', n)
\end{smallmatrix}}\hspace{-4mm}{{{\bm{h}}_{k,m',n}^{\rm{H}}}{{\bm{w}}_{k',{m}',n}}{{d}_{{k}',{m}',n}}}+\\
z_{k,m,n},
\end{equation}
%\vspace{-1mm}
%\begin{equation}
%\label{optz1}
%\begin{aligned}
%& {{y}_{k,m,n}}={{\bm{h}}_{k,m,n}^{\rm{H}}}{{\bm{w}}_{k,m,n}}{{d}_{k,m,n}}+\sum\limits_{\begin{smallmatrix}
% {m}'\in \mathcal{M}\backslash m \\
% {k}'=f(m', n)
%\end{smallmatrix}}{{{\bm{h}}_{k,m',n}^{\rm{H}}}{{\bm{w}}_{k',{m}',n}}{{d}_{{k}',{m}',n}}}\\
%& \text{~~~~~~~~}+ z_{k,m,n},
%\end{aligned}
%\end{equation}
where $\text{ }y_{k,m,n}\in \mathbb{C}$ denotes the received symbol. $\bm{h}_{k,m,n}^{\rm{H}}\in \mathbb{C}^{{1\times N_\mathrm{Tx}}} $ is the complex channel vector between BS $m$ and user $k$, and let $\bm{w}_{k,m,n}\in \mathbb{C}^{{N_\mathrm{Tx}}\times 1} $ be the associated beamforming vector. $z_{k,m,n}\sim{\mathcal{CN}(0,1)}$ is the background noise at user $k$ and $d_{k,m,n}$ is the normalized transmitted symbol from BS $m$ to user $k$.
%\sum\limits_{m=1}^{M}\sum\limits_{k=1}^{K}                (i.e., a user decodes its intended signals by treating all interfering signals as noise)
%\subsection{Weighted Sum-rate Maximization (WSRMax) Problem }

The signal-to-interference-plus-noise ratio (SINR) of the $k$th user from cell $m$ scheduled on subcarrier $n$ is given by
\vspace{0mm}
\begin{equation}
\label{optz2}
{{\gamma }_{k,m,n}}({{\bm{w}}_{k,m,n}})=\frac{|{{\bm{h}}_{k,m,n}^{\rm{H}}}{{\bm{w}}_{k,m,n}}|^2}{1+\sum\nolimits_{\begin{smallmatrix}
 {m}'\in \mathcal{M}\backslash m \\
  {k}'=f(m', n)
 \end{smallmatrix}}{|{{\bm{h}}_{k,m',n}^{\rm{H}}}{{\bm{w}}_{k',{m}',n}}}|^2}.
\end{equation}
%\vspace{-.5mm}
The instantaneous rate achieved by user $k$ from cell $m$ on subcarrier $n$ is given by ${{r}_{k,m,n}}={{\log }_{2}}(1+{{\gamma }_{k,m,n}}({{\bm{w}}_{k,m,n}}))$. Positive scalar $\alpha_{k,m}$ is the weight associated with user $k$ in cell $m$. The WSRMax problem under BS transmit power constraint is 
%The instantaneous rate for user $k$ is given by ${{C}_{k,m}}=\sum\nolimits_{n\in {{\mathcal{N}}_{k,m}}}{{{r}_{k,m,n}}}$, where ${{r}_{k,m,n}}={{\log }_{2}}(1+{{\gamma }_{k,m,n}})$ is the rate achieved by user $k$ from cell $m$ on subcarrier $n$, and the set ${\mathcal{N}_{k,m}}$ is defined as ${{\mathcal{N}}_{k,m}}:=\left\{ n\text{~} |\text{~} k=f(m,n) \right\}$. Positive scalar $\alpha_{k,m}$ is the weight associated with user $k$ in cell $m$. Then the nonconvex WSRMax problem under per-BS power constraint has the form
\vspace{-1mm}
\begin{equation}
\label{optz4}
\begin{aligned}
& \underset{\left\{{{\bm{w}}_{k,m,n}}\right\}}{\mathop{\mathrm{max} }}\,\text{~~~~}\sum\nolimits_{m=1}^{M}{\sum\nolimits_{\begin{smallmatrix}
 n=1 \\
 k=f(m, n)
\end{smallmatrix}}^{N}{{{\alpha }_{k,m,n}}}{{\log }_{2}}(1+{{\gamma }_{k,m,n}}({{\bm{w}}_{k,m,n}}))}\\
& \text{s.t.}\text{~~~~~~~~} \sum\nolimits_{\begin{smallmatrix}
 n=1 \end{smallmatrix}}^{N}{||{{\bm{w}}_{k,m,n}}|{{|}_2^{2}}\le {{P}_{\max ,m}}},\text{  } m=1,...,M,
\end{aligned}
\end{equation}
%\begin{equation}
%\label{optz3}
%\begin{aligned}
%& \underset{{{\bm{w}}_{k,m,n}}}{\mathop{\mathrm{max} }}\,\text{~~~~}\sum\nolimits_{\forall m}\sum\nolimits_{\forall k}{{{\alpha}_{k,m}}{{C}_{k,m}}}\\
%& \text{s.t.}\text{~~~~~~} \sum\nolimits_{\begin{smallmatrix}
% n=1 \end{smallmatrix}}^{N}{||{{\bm{w}}_{k,m,n}}|{{|}_2^{2}}\le {{P}_{\max, m}}},\text{  } m=1,...,M
%\end{aligned}
%\end{equation}
%\vspace{-1mm}
where ${\alpha }_{k,m,n}={\alpha }_{k,m}, \forall n$ and $||\cdot||_2$ refers to $l_2$ norm. $P_{\max,m }$ is the transmit power constraint of BS $m$. The WSRMax problem in \eqref{optz4} is NP-hard, therefore, very difficult to solve. Herein, we discuss our provably fast convergent convex approximated iterative WSRMax algorithm for solving this nonconvex problem.
%where ${\alpha }_{k,m,n}={\alpha }_{k,m}, \forall n$ and $||\cdot||_2$ refers to $l_2$ norm. $\mathcal{W}:=\left\{{\bm{w}}_{k,m,n};\text{ }m\in M,\text{ } n\in N\right\}$ is the set of all beamforming vectors, and $P_{\max,m }$ is the transmit power constraint of BS $m$. Let $\mathcal{W}_m$ be the set of beamformers for cell $m$. Herein, we discuss our proposed fast convergent iterative WSRMax algorithm for solving this problem.

\vspace{0mm}
%================================================
\section{Low Complexity SPCA Beamformer Design}
\label{section:chapter4-3}
%================================================
\vspace{0mm}
%To transform the nonconvex WSRMax problem into a solvable convex form, we begin with expressing the problem \eqref{optz3} into a standard form as 
%\vspace{-2mm}
%\begin{equation}
%\label{optz4}
%\begin{aligned}
%& \underset{{{\bm{w}}_{k,m,n}}}{\mathop{\mathrm{max} }}\,\text{~~~~}\sum\nolimits_{m=1}^{M}{\sum\nolimits_{\begin{smallmatrix}
% n=1 \\
% k=f(m, n)
%\end{smallmatrix}}^{N}{{{\alpha }_{k,m,n}}}{{\log }_{2}}(1+{{\gamma }_{k,m,n}}({{\bm{w}}_{k,m,n}}))}\\
%& \text{s.t.}\text{~~~~~~} \sum\nolimits_{\begin{smallmatrix}
% n=1 \end{smallmatrix}}^{N}{||{{\bm{w}}_{k,m,n}}|{{|}_2^{2}}\le {{P}_{\max ,m}}},\text{  } m=1,...,M
%\end{aligned}
%\end{equation}
%\vspace{0mm}
To transform the nonconvex WSRMax problem into a solvable convex form, we begin with introducing a set of indices $L$ as 
 $L:=\left\{\left\{k,m,n\right\},\forall m,n\text{  }|\text{  }k=f(m,n)\right\}$. Without loss of generality, the subsets of $L$ can be expressed as $L:=\bigg\{\{k,1,1\},\cdots,\{k,1,N\},\{k,2,1\},\cdots,\{k,2,N\},\cdots,\{k,M,1\},\cdots$, $\{k,M,N\}\bigg\}$. Under the considered system model and subcarrier allocation scheme, the objective function in \eqref{optz4} is a function of $T=MN$ optimization variables, and can be expressed as $ \underset{\left\{{{\bm{w}}_{L_t}}\right\}}{\mathop{\mathrm{max} }}\,\sum\nolimits_{t=1}^{T}{{{\alpha }_{L_t}}{{\log }_{2}}(1+{{\gamma }_{L_t}}({{\bm{w}}_{L_t}}))}={{\log }_{2}} \left(\underset{\left\{{{\bm{w}}_{L_t}}\right\}}{\mathop{\mathrm{max} }}\,\prod\nolimits_{t=1}^{T}{{{(1+{{\gamma }_{L_t}}({{\bm{w}}_{L_t}}))}^{{{\alpha }_{L_t}}}}}\right),$
%\begin{equation}
%\label{optz5}
% \underset{\bm{\mathcal{W}}}{\mathop{\mathrm{max} }}\,\sum\limits_{t=1}^{T}{{{\alpha }_{L_t}}{{\log }_{2}}(1+{{\gamma }_{L_t}})}={{\log }_{2}} \left(\underset{\bm{\mathcal{W}}}{\mathop{\mathrm{max} }}\,\prod\limits_{t=1}^{T}{{{(1+{{\gamma }_{L_t}})}^{{{\alpha }_{L_t}}}}}\right),
%\end{equation}
% For example, let $t=69$, $M=3$ and $N=64$. Under the current structure of the set $L$, we get $m=2$, $n=5$. Therefore, $L_{69}$ denotes the index set $\{k, m, n\}=\{k, 2, 5\}$.
where $L_t$ denotes the $t$th subset in $L$. The corresponding indices in the $t$th subset are defined as  $\{k,m,n\}=\Big\{k, \lceil{t/N}\rceil, t-\big((\lceil{t/N}\rceil-1)N\big)\Big\}$, where $\lceil \cdot \rceil$ defines the ceiling operation. Now, following the monotonicity of logarithmic function, the log function in the subsequent optimization problems can be eliminated, and introducing new slack variable $r_{L_t}$, we can recast \eqref{optz4} as
\text{}$\underset{\left\{{{\bm{w}}_{L_t}},{{r}_{L_t}}({{\bm{w}}_{L_t}})\right\}}{\mathop{\mathrm{max} }}$\text{~~}$\prod\limits_{t=1}^{T}{r}_{L_t}^{{\alpha}_{L_t}}({{\bm{w}}_{L_t}})$\\
\text{~~~s.t. }\text{~~~~~~~}C1: $\sum\limits_{\begin{smallmatrix}
 n=1  \end{smallmatrix}}^{N}{||{{\bm{w}}_{k,m,n}}|{{|}_2^{2}}\le {{P}_{\max ,m}}},\text{  } m=1,...,M$\text{~~~~~~~~~}(3a)\\
\text{~~~~~~~~~~~~}\text{~~}C2: $r_{L_t}({{\bm{w}}_{L_t}})\le {{\gamma }_{L_t}}({{\bm{w}}_{L_t}})+1,\text{~~}\forall L_t\in L, \text{~}t=1,...,T$,
%\begin{equation}
%\label{optz6}
%\begin{aligned}
%&  \underset{\bm{\mathcal{W}},{{r}_{L_t}}}{\mathop{\mathrm{max} }}\,\text{~~~~~~}\prod\limits_{t=1}^{T}{r}_{L_t}^{{\alpha}_{L_t}}\\
%& \text{subject to}\text{~} \text{C1}:\sum\limits_{\begin{smallmatrix}
% n=1  \end{smallmatrix}}^{N}{||{{\bm{w}}_{kmn}}|{{|}_2^{2}}\le {{P}_{\max ,m}}},\text{  } m=1,...,M\\
%&\text{~~~~~~~~~~~~}\text{C2}:r_{L_t}\le {{\gamma }_{L_t}}+1,\text{~~}\forall L_t\in L, \text{~}t=1,...,T
%\end{aligned}
%\end{equation}

The equivalence between \eqref{optz4} and (3a) is recognized by the fact that all the constraints in C2 are active at the optimum. Contrarily, a strictly larger objective can be obtained by increasing ${r}_{L_t}({{\bm{w}}_{L_t}})$ without violating the constraints\cite{Mirza}. Let $\mathcal{W}_m$ be the set of beamformers for cell $m$. Using $\mathrm{vec}(\cdot)\footnote{$\mathrm{vec}(\cdot)$/ $\mathrm{diag}(\cdot)$ stacks all/diagonal elements of a matrix into a column vector.}$, power constraints in C1 can be simplified as $|| \mathrm{vec}\left( {{\mathcal{W}}_{m}} \right)|{{|}_{2}}\le \sqrt{{{P}_{\max, m}}}$.
%\begin{equation}
%\label{optz7}
%|| \mathrm{vec}\left( {{\mathcal{W}}_{m}} \right)|{{|}_{2}}\le \sqrt{{{P}_{\max, m}}}.
%\end{equation}
Further, by introducing auxiliary variables $\beta_{L_t}\ge 0$, the WSRMax problem can be formulated as 
\begin{equation}
\label{optz8}
\begin{aligned} 
&  \underset{\left\{{{\bm{w}}_{L_t,}} {{r}_{L_t}}({{\bm{w}}_{L_t}}), \beta_{L_t}\right\}}{\mathop{\mathrm{max} }}\,\text{~}\prod\nolimits_{t=1}^{T}{r}_{L_t}^{{\alpha}_{L_t}}({{\bm{w}}_{L_t}})\\
& \text{s.t.}\text{~~~~~~~~~} \text{C1}:|| \mathrm{vec}\left( {{\mathcal{W}}_{m}} \right)|{{|}_{2}}\le \sqrt{{{P}_{\max, m}}}, \\
&\text{~~~~~~~~~~~~}\text{C2}:\sqrt{{1+\sum\nolimits_{\begin{smallmatrix}
 {m}'\in \mathcal{M}\backslash m \\
  {k}'=f(m', n)
 \end{smallmatrix}}{|{{\bm{h}}_{k,m',n}^{\rm{H}}}{{\bm{w}}_{k',{m}',n}}}|^2}}\le {{\beta }_{{{L}_{t}}}},\\
&\text{~~~~~~~~~~~~}\text{C3}:{{\beta}_{{{L}_{t}}}}(r_{L_t}({{\bm{w}}_{L_t}})-1)^{1/2}\le {{\bm{h}}_{L_t}^{\rm{H}}}{{\bm{w}}_{L_t}}.\\
&\text{~~~~~~~~~~~~}\text{C4}\footnotemark:\operatorname{\Im}\{{{\bm{h}}_{L_t}^{\rm{H}}}{{\bm{w}}_{L_t}}\}=0,\text{ } \Im\left\{\cdot\right\}=\text{Imaginary part.}\\
\end{aligned}
\end{equation}
\footnotetext{For any $\theta$, we have $|\bm{h}_{L_t}^{\rm{H}}\bm{w}_{L_t}|^2=|\bm{h}_{L_t}^{\rm{H}}\bm{w}_{L_t}e^{j\theta}|^2$. Therefore, by choosing $\theta$ such that $\Im\{{{\bm{h}}_{L_t}^{\rm{H}}}{{\bm{w}}_{L_t}}\}=0$ does not affect the optimality of \eqref{optz8}. } 
Let $\bm{W}_{\operatorname{int}}\in\mathbb{C}^{N_\mathrm{Tx} \times(M-1)}$ and $\bm{H}_{\operatorname{int}}\in\mathbb{C}^{{(M-1)\times N_\mathrm{Tx}}} $ be the aggregated beamforming and channel matrices, respectively, containing beamformers and the channels from all interfering BSs corresponding to the constraint C2 of \eqref{optz8}. Therefore, we can reformulate the constraint C2 equivalently as 
\begin{equation}
\label{tripX}
{{\left\| {{\left[ \begin{matrix}1 & \mathrm{diag}\left( {{\bm{H}}^{{\rm{H}}}_{\operatorname{int}}}{{\bm{W}}_{\operatorname{int}}} \right)  \\ \end{matrix} \right]}^{\mathrm{T}}} \right\|}_{2}}\le {{\beta }_{{{L}_{t}}}}.
\end{equation}
%\begin{equation}
%\label{optz9}
%{{\left\| {{\left[ \begin{matrix}1 & \mathrm{diag}\left( {{\bm{H}}_{\operatorname{int}}}{{\bm{W}}_{\operatorname{int}}} \right)  \\ \end{matrix} \right]}^{\mathrm{Tx}}} \right\|}_{2}}\le {{\beta }_{{{L}_{t}}}}.
%\end{equation}
One can notice that most of the constraints of problem \eqref{optz8} are convex, except the constraints in C3. 
%Therefore, the constraints C1 and C2 do not require any approximation.
%, and we can concentrate on the following nonconvex problem only
%\begin{equation}
%\label{optz10}
%\begin{aligned} 
%&{\mathop{\mathrm{max} }}\text{~~}\prod\limits_{t=1}^{T}{{{r}_{L_t}}}\\
%& \text{subject to}\text{~~} {{\beta}_{{{L}_{t}}}}(r_{L_t}^{1/{{\alpha}_{L_t}}}-1)^{1/2}\leq & {{\bm{h}}_{L_t}}{{\bm{w}}_{L_t}}.\\
%\end{aligned}
%\end{equation}
%\noindent By introducing auxiliary variables $x_{L_t}:=r_{L_t}^{1/{{\alpha}_{L_t}}}$ and rearranging the terms, we can further reformulate problem \eqref{optz10} as
%\begin{equation}
%\label{optz11}
%\begin{aligned} 
%&{\mathop{\mathrm{max} }}\text{~~}\prod\limits_{t=1}^{T}x_{L_t}^{{\alpha}_{L_t}}\\
%& \text{subject to}\text{~~} {{\beta}_{{{L}_{t}}}}(x_{L_t}-1)^{1/2}\leq & {{\bm{h}}_{L_t}}{{\bm{w}}_{L_t}}.\\
%\end{aligned}
%\end{equation}
Clearly, the function $\frac{1}{2}\left(\frac{r_{L_t}({{\bm{w}}_{L_t}})-1}{\phi_{L_t}}+\phi_{L_t} \beta_{L_t}^{2}\right)$ upper bounds the constraint for all positive $\phi_{L_t}$s (i.e, solution of the approximated problem will be a feasible point of the original problem)\cite{Beck}, arising from the inequality between arithmetic and geometric means of $\frac{r_{L_t}({{\bm{w}}_{L_t}})-1}{\phi_{L_t}}$ and $\phi_{L_t} \beta_{L_t}^{2}$. Indeed, it follows from this observation $\left(\sqrt{r_{L_t}({{\bm{w}}_{L_t}})-1}-\phi_{L_t}\beta_{L_t}\right)^{2}\geq 0$.
%Note that the WSRMax problem in \eqref{optz8} with the convex approximated constraints C3 as $\frac{1}{2}\left(\frac{r_{L_t}({{\bm{w}}_{L_t}})-1}{\phi_{L_t}}+\phi_{L_t} \beta_{L_t}^{2}\right)\leq {{\bm{h}}_{L_t}^{\rm{H}}}{{\bm{w}}_{L_t}}$
%is still nonconvex due to the nonconvex objective function. 

Note that the optimization function $\underset{{{\bm{w}}_{L_t,}}{{r}_{L_t}}, {{\beta }_{L_t}}}{\mathop{\mathrm{maximize} }}\,\text{~~} \prod\nolimits_{t=1}^{T}{r_{L_t}^{{{\alpha }_{L_t}}}}$ in \eqref{optz8} is nonconvex in its current form. Logarithmic variable transformation: $s_{L_t}({{\bm{w}}_{L_t}}):=\log(r_{L_t}({{\bm{w}}_{L_t}}))$ produces the equivalent convex optimization problem as  
\begin{equation}
\label{optz14}
\begin{aligned} 
&\underset{\left\{{{\bm{w}}_{L_t,}}, \beta_{L_t}, {{s}_{L_t}({{\bm{w}}_{L_t}})}\right\}} {\mathop{\mathrm{max} }}\text{~~~}\exp\left(\sum\nolimits_{t=1}^{T}s_{L_t}({{\bm{w}}_{L_t}}){{\alpha}_{L_t}}\right)\\
& \text{s.t.}\text{~~~~C3:}\text{~} \frac{1}{2}\left(\frac{\exp (s_{L_t}({{\bm{w}}_{L_t}}){{\alpha}_{L_t}})-1}{\phi_{L_t}}+\phi_{L_t} \beta_{L_t}^{2}\right)\leq & {{\bm{h}}_{L_t}^{\rm{H}}}{{\bm{w}}_{L_t}}.\\ 
\end{aligned}
\end{equation}
%$\frac{1}{2}\left(\frac{{ (z_{k}}-1)}{\phi_{k}^{(j)}}+\phi_{k}^{(j)} \delta_{k}^{2}\right)\leq\bm{h}_{{{m}_{k}},k}^{\rm{H}}{{\bm{f}}_{k}}$
%Since the constraints C1 and C2 of \eqref{optz8} are not functions of ${{\alpha}_{L_t}}$ and $r_{L_t}$, the change of variables does not affect their convexities. The convex problem \eqref{optz14} can be efficiently solved \cite{Boyd}. 
%Interestingly, GP problem in \eqref{optz13} automatically transforms into a SOCP problem with equivalent SOCP form for constraint C3 if the weights on subcarriers for all the users are equal or equal to one. This is because maximizing $\prod\limits_{t=1}^{T}r_{L_t}^{{\alpha}_{L_t}}$ with ${{\alpha}_{L_1}}={{\alpha}_{L_2}}=\cdots={{\alpha}_{L_T}}\hspace{1mm} (=1)$ is equivalent to the maximization of the product of the optimization variables and $\prod\limits_{t=1}^{T}r_{L_t}$ can be easily expressed as hyperbolic, i.e., SOCP constraints.
%and an additional constraint\footnote{For any $\theta$, we have $|\bm{h}_{L_t}\bm{w}_{L_t}|^2=|\bm{h}_{L_t}\bm{w}_{L_t}e^{j\theta}|^2$. Therefore, by choosing $\theta$ such that $\Im\{{{\bm{h}}_{L_t}}{{\bm{w}}_{L_t}}\}=0$ does not affect the optimality of \eqref{optz13}.} $\Im{({{\bm{h}}_{L_t}}{{\bm{w}}_{L_t}})}=0$
Here, the objective function $\exp\left(\sum\nolimits_{t=1}^{T}s_{L_t}({{\bm{w}}_{L_t}}){{\alpha}_{L_t}}\right)$ is logarithmically convex. This is because the function $g(x)=e^{f(x)}$ is logarithmically convex if $f(x)$ is a convex function. In this case $f(x)=\sum\nolimits_{t=1}^{T}s_{L_t}({{\bm{w}}_{L_t}}){{\alpha}_{L_t}}$, clearly a convex function. Hereafter, we express $s_{L_t}({{\bm{w}}_{L_t}})$ and $r_{L_t}({{\bm{w}}_{L_t}})$ with $s_{L_t}$ and $r_{L_t}$, respectively and employ the following procedure to solve \eqref{optz14}.\newline
\noindent \textbf{Iteration 1:} Choose the first feasible point $(s_{L_t}^{(0)},\beta_{L_t}^{(0)})$. Compute $\phi_{L_t}^{(1)}={\sqrt{{\exp (s_{L_t}^{(0)}{{\alpha}_{L_t}})}-1}}/{\beta_{L_t}^{(0)}}$. This choice guarantees the equality between the original function and its approximations as well as equality between their gradients at the point $(s_{L_t}^{(0)},\beta_{L_t}^{(0)})$. The gradients equality guarantees that the Karush-Kuhn-Tucker (KKT) conditions of the original nonconvex problem will be satisfied as the approximated solution converges. Solve the problem in \eqref{optz14} with C3 as $\frac{1}{2}\left(\frac{{\exp (s_{L_t}{{\alpha}_{L_t}})}-1}{\phi_{L_t}^{(1)}}+\phi_{L_t}^{(1)} \beta_{L_t}^{2}\right)\leq{{\bm{h}}_{L_t}^{\rm{H}}}{{\bm{w}}_{L_t}}$, and denote it by $(s_{L_t}^{(1)},\beta_{L_t}^{(1)})$.\newline
\noindent \textbf{Iteration $i$:} For a given point $(s_{L_t}^{(i-1)},\beta_{L_t}^{(i-1)})$ (the solution of the previous convex problem) compute $\phi_{L_t}^{(i)}={\sqrt{{\exp (s_{L_t}^{(i-1)}{{\alpha}_{L_t}})}-1}}/{\beta_{L_t}^{(i-1)}}$ and solve \eqref{optz14} with C3 as $\frac{1}{2}\left(\frac{{\exp (s_{L_t}{{\alpha}_{L_t}})}-1}{\phi_{L_t}^{(i)}}+\phi_{L_t}^{(i)} \beta_{L_t}^{2}\right)\leq{{\bm{h}}_{L_t}^{\rm{H}}}{{\bm{w}}_{L_t}}$. Continue with this procedure until a convergence point is reached. 
%The same procedures can also be followed for solving the GP problem \eqref{optz13} with appropriate changes in the initialization variables and setting $\phi_{L_t}^{(1)}$ as $\sqrt{{ (r_{L_t}^{(0)}-1)}}/\beta_{L_t}^{(0)}$.
\vspace{0mm}
\begin{algorithm}
  \SetAlgoLined
%  \KwData{this text}
%  \KwResult{how to write algorithm with \LaTeX2e }
  \textbf{Initialization}: Tolerance $\epsilon >0$,  feasible points $(s_{L_t}^{(0)}, \beta_{L_t}^{(0)})$, $\phi_{L_t}^{(1)}={\sqrt{{\exp (s_{L_t}^{(0)}{{\alpha}_{L_t}})}-1}}/{\beta_{L_t}^{(0)}}$,\
  $i=1$, $N_{\rm{iter}}=20; $
  \BlankLine
  \While{$i< N_{\rm{iter}}$}{
   \textbf{Solve: }$\underset{\left\{{{\bm{w}}_{L_t}}, \beta_{L_t}, {{s}_{L_t}}\right\}}{\mathop{\mathrm{max} }}\text{~~~~~~~}\exp\left(\sum\nolimits_{t=1}^{T}s_{L_t}{{\alpha}_{L_t}}\right)$\;\vspace{1mm}
     \text{~~~~~~~subject to }\text{~}C1, C2  and C4 of  \eqref{optz8}\;\vspace{1mm}
       \text{~~~~~~~~~~~~~~~~~~~~}C3:$\frac{1}{2}\left(\frac{{\exp (s_{L_t}{{\alpha}_{L_t}})}-1}{\phi_{L_t}^{(i)}}+\phi_{L_t}^{(i)} \beta_{L_t}^{2}\right)\leq{{\bm{h}}_{L_t}^{\rm{H}}}{{\bm{w}}_{L_t}}$\;\vspace{1mm}
       Denote the solution as $s_{L_t}^{(*)}$, $\beta_{L_t}^{(*)}.$\;\vspace{2mm}
                          \eIf{$|\exp\left(\sum\limits_{t=1}^{T}s_{L_t}^{(*)}{{\alpha}_{L_t}}\right)-\exp\left(\sum\limits_{t=1}^{T}s_{L_t}^{(i)}{{\alpha}_{L_t}}\right)|\le \epsilon$}{
      ${\textbf{break}}$\;
      }{
      $i=i+1$, \{$s_{L_t}^{(i-1)}=s_{L_t}^{(*)}$, $\beta_{L_t}^{(i-1)}=\beta_{L_t}^{(*)}$\}\;
      $\phi_{L_t}^{(i)}={\sqrt{\exp (s_{L_t}^{(i-1)}{{\alpha}_{L_t}})-1}}/{\beta_{L_t}^{(i-1)}}$\}\;
      }
    }
  \caption{SPCA-Exp algorithm for WSRMax}
   \label{algo}
\end{algorithm}
This is an exponential optimization problem, and we designate this proposed solution as SPCA-Exp. 

Furthermore, the geometric-mean is a concave function for nonnegative affine optimization variables, and admits SOCP formulation \cite{Alizadeh}. Maximization of $\prod\nolimits_{k=1}^{K}{r_{L_t}^{{{\alpha }_{k}}}}({{\bm{w}}_{L_t}})$ can be equivalently expressed as
\begin{equation}
\label{optz15}
\begin{aligned} 
&\underset{\left\{{{\bm{w}}_{L_t,}}{{r}_{L_t}}, {{\beta }_{L_t}}\right\}}{\mathop{\mathrm{maximize} }}\,\hspace{2mm}{{\left( \prod\nolimits_{t=1}^{T}{r_{L_t}^{{{\alpha }_{L_t}}}}({{\bm{w}}_{L_t}})\right)}^{\frac{1}{\sum\nolimits_{t=1}^{T}{{{\alpha }_{L_t}}}}}}\\
& \text{s.t.}\text{~~~~C3:}\text{~} \frac{1}{2}\left(\frac{{ (r_{L_t}}({{\bm{w}}_{L_t}})-1)}{\phi_{L_t}}+\phi_{L_t} \beta_{L_t}^{2}\right)\leq\bm{h}_{L_t}^{\rm{H}}{{\bm{w}}_{L_t}}.\\ 
\end{aligned}
\end{equation}
which is a weighted geometric-mean (WGM) optimization problem, therefore, can be equivalently cast as SOCP. We denote this approach as SPCA-WGM solution. Consequently, \eqref{optz8} can be transformed into an SOCP if all $\alpha_{L_t}$ are rational numbers \cite{Alizadeh}. For a special case when $\alpha_{L_t}=1$ for all $L_t$, the function ${\mathop{\mathrm{max} }}\text{~~}\prod\limits_{t=1}^{T}r_{L_t}^{{\alpha}_{L_t}}({{\bm{w}}_{L_t}})$ becomes equivalent to the maximization of the geometric-mean of the optimization variables, i.e., $\underset{{{\bm{w}}_{L_t}},{{r}_{L_t}}({{\bm{w}}_{L_t}}),\beta_{L_t}}{\mathop{\mathrm{max} }}\,\prod\limits_{t=1}^{T}{{{r}_{L_t}({{\bm{w}}_{L_t}})}}:\Leftrightarrow \underset{{{\bm{w}}_{L_t}},{{r}_{L_t}}({{\bm{w}}_{L_t}}),\beta_{L_t}}{\mathop{\mathrm{max} }}\,\prod\limits_{t=1}^{T}({{r}_{L_t}}({{\bm{w}}_{L_t}}))^{1/T}$. Therefore, \eqref{optz8} has an equivalent SOCP formulation for all practical purposes. 
%We can follow similar procedure to solve the WSRMax problem in \eqref{optz15}.

Therefore, this algorithm solves the convex approximated problem \eqref{optz14} in an iterative manner and repeats until convergence. Initial guesses of ${{\phi }_{{{L}_{t}}}}$s are very crucial to the feasibility of the successive optimization method. The randomly generated ${{\phi }_{{{L}_{t}}}}$s would lead to infeasible solution. The other numerical issue that requires attention is that if one of the $s_{L_t}$s goes to zero, the algorithm faces the problem of `dividing by zero' since it needs to calculate $1/{\phi_{L_t}}$. Initial guess of ${{\phi }_{{{L}_{t}}}}$s are very crucial to the feasibility of the successive optimization method. The randomly generated ${{\phi }_{{{L}_{t}}}}$s would lead to infeasible solution. To guarantee the feasibility at the first step, we follow the steps in \textbf{Procedure}  to find $\phi_{L_t}^1$s.
\vspace{-7mm}
\begin{center}
\line(1,0){240}
\end{center}
\vspace{-3mm}
$\textbf{Procedure:} \text{ For generating good initial values of  } \phi_{L_t}^{(1)}$
\vspace{-5mm}
\begin{center}
\nointerlineskip
\line(1,0){240}
\end{center}
\vspace{-3mm}
$ {\textbf{Step 1: }}\text{Channel-matched beamforming vectors are obtained} $\\
$\text{such that BS power constraint is satisfied for all cells, i.e.,} $\\
$ \text{ }{{\bm{w}}_{kmn}}=\sqrt{{{{P}_{\max,m }}}/{N}}({{{\bm{h}}_{kmn}}}/{||{{\bm{h}}_{kmn}}||_2}), \forall m,n \text{ and }k=f(m,n).$ \\
$ {\textbf{Step 2: }}\text{ Choose } {\beta }^{(0)}_{{L}_{t}} \text{ to be the value that gives the equality }$\\
$ \text{in \eqref{tripX}. } $\\
$ {\textbf{Step 3: }}\text{Obtain the value of }r_{L_{t}}\text{ from C3 of \eqref{optz8} according to}$\\
$ \text{the obtained value of } {\beta }^{(0)}_{{L}_{t}}\text{ using equality with } |{{\bm{h}}_{L_t}^{\rm{H}}}{{\bm{w}}_{L_t}}|. $\\
$ {\textbf{Step 4: }}\text{Compute corresponding value of }x_{L_t}^{(0)}=r_{L_{t}}^{1/\alpha_{L_t}}\text{ and }{{\phi }_{{{L}_{t}}}}  $\\
$ \text{values are calculated as }\phi_{L_t}^{(1)}=\sqrt{{\exp (s_{L_t}^{(0)}{{\alpha}_{L_t}})-1}}/\beta_{L_t}^{(0)}.$\\
\vspace{-8mm}
\begin{center}
\nointerlineskip
\line(1,0){240}
\end{center}
%\end{aligned} \] 
% follow the steps in \textbf{Procedure}  to find $\phi_{L_t}^1$s.  (please see C3 in \eqref{optz14} and (8) in \cite{Tran}). We can easily overcome this numerical instability problem by constraining $s_{L_t}$ as $s_{L_t}\ge \eta$, where $\eta$ is a very small positive quantity, without affecting much the achieved sum-rate.

Now, we analyze the convergence behavior of our proposed WSRMax algorithm. Let us consider the $(i+1)$th iteration of our optimization process. If we replace $(s_{L_t}, \beta_{L_t})$ with $(s_{L_t}^{(i)}, \beta_{L_t}^{(i)})$ and ${\bm{w}_{L_t}}$ by ${\bm{w}_{L_t}^{(i)}}$, constraints C1-C4 are still satisfied. Therefore, the optimal solution in the $i$th iteration is a feasible point of the optimization problem in the $(i+1)$th iteration. Thus, the objective value in the $(i+1)$th iteration is larger than or equal to that in the $i$th iteration. Therefore, the WSRMax algorithm yields nondecreasing sequence of objective values.

 In general, calculation of the beamformers requires CSI at the BS. Also, the beamformers need to be updated when the users move to a different location. The frequency of CSI update depends on many factors such as feedback budget, fading severity, channel correlation, Doppler frequency etc. In all practical beamforming systems, the signal processor should output a result within a short time (after CSI is available), which is usually several transmission time intervals (TTI). In order to be able to adapt quickly to the changing conditions in the radio link, a communication system must have shorter TTIs. Furthermore, CSI feedback budget also affects the CSI update frequency. However, in this letter, to keep the analysis tractable and simple, we consider perfect CSI and the proposed algorithm should be run as soon as the CSI is available. Furthermore, due to channel estimation error, quantization error, and delay in feedback, the CSI update may be erroneous, and performance of the system degrades because of mismatched beamforming.
\vspace{0mm}
\section{Numerical Results}
\label{section:chapter4-4}
%===================================
In this section, we numerically analyze the performance of the proposed algorithm on a cellular network with 1-cell frequency reuse factor. The distance between neighboring BSs is 1000 m. The frequency-selective channel coefficients over $N=$64 subcarriers are modeled as ${{\bm{h}}_{k,m,n}}={{\left(200\frac{1}{{{l}_{k,m}}}\right)}^{3.5}}{{\Phi }_{k,m,n}}{{\Lambda }_{k,m,n}}$, where ${l}_{k,m}$ is the distance between BS $m$ and user $k$. $10{{\log }_{10}}({{\Phi }_{k,m,n}})$ accounts for  lognormal shadowing and is distributed as $\mathcal{RN}(0,8)$, and ${{\Lambda }_{k,m,n}}\sim\mathcal{CN}(0,1)$ accounts for Rayleigh fading. All the BSs are subjected to the equal transmit power constraints, i.e., ${{P}_{\max,m}}={{P}_{\max }},\forall m$. We also consider that perfect channel state information (CSI) is available both at the BSs and users. The user scheduling is performed randomly. We use the disciplined convex optimization toolbox CVX \cite{CVX} with internal solver SeDuMi \cite{Strum}. Note that CVX invoke its successive approximation method for SPCA-Exp approach in both the objective and  the constraints.
%\footnote{A predetermined order or sequence is not followed while allocating subcarriers to the users. Also note that the non-overlapping subcarrier scheduling in each cell does not restrict the adaptability of the proposed SPCA-WSRMax algorithm for multiple active users in one subcarrier in one cell.}
%-------------------------------------------------------------------------------------------- 
%\vspace{-2mm}
 \begin{figure}
  \centering
   \includegraphics[scale=0.18]{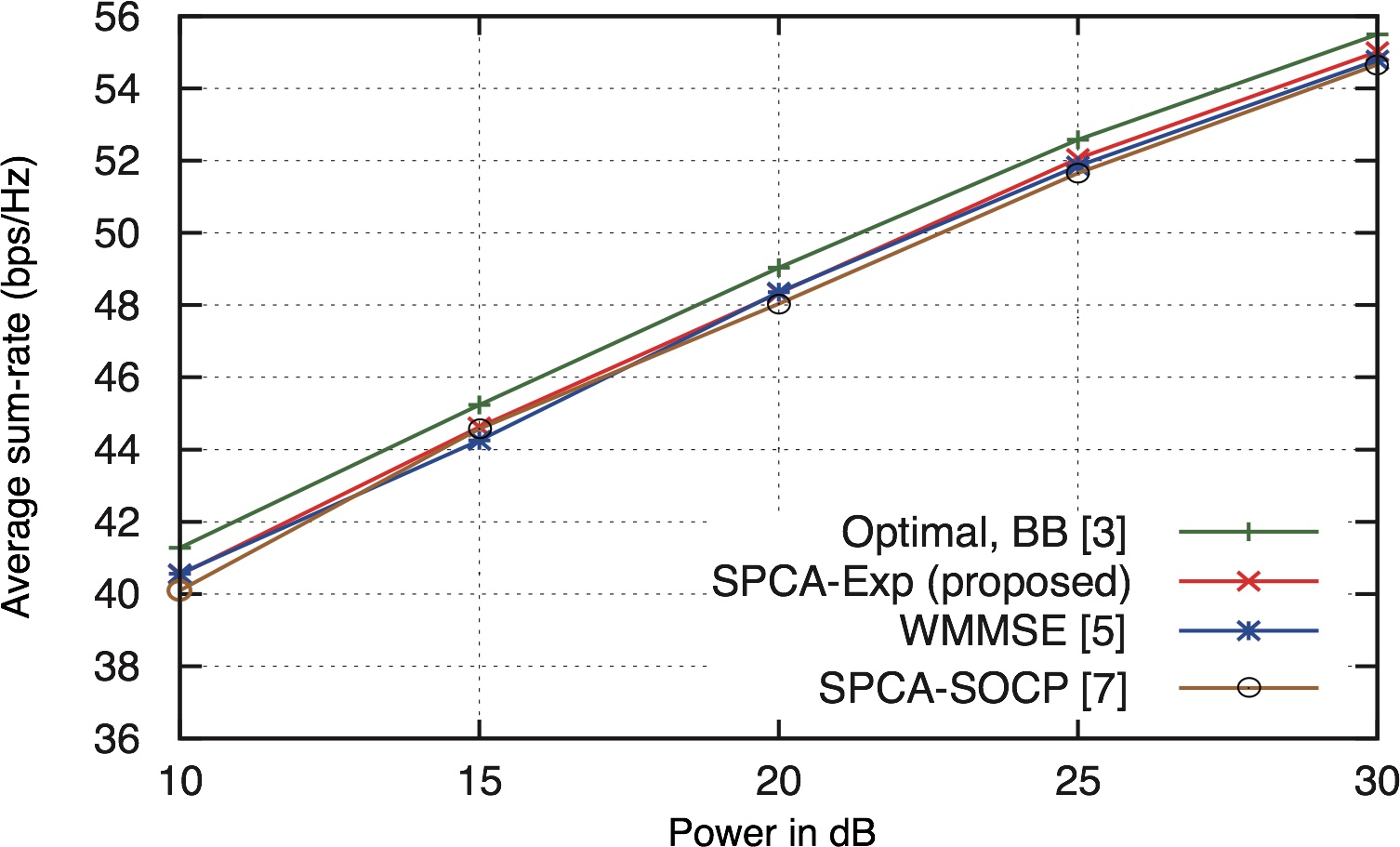}
   \caption{Average sum-rate performances for different WSRMax algorithms.}
   \label{BER1}
 \end{figure}
 %-----------------------------------------------------------------------------------------------
%AM algorithm in \cite{Wang}, soft interference nulling (SIN) approach in \cite{Chris}
%\vspace{-2mm}

In Fig.~\ref{BER1}, we depict the average sum-rate ($\alpha_{k,m}$=1) as a function of total transmit power per-BS with $M=3$, $N_\mathrm{Tx}=2$ and $K=2$. The average sum-rate achieved by our proposed algorithm is compared to WMMSE based power minimization, SPCA-SOCP scheme, and global optimal design using BB method. The gap tolerance between the lower and upper bounds is set to 0.01. With this gap tolerance, the BB method converges to a sum-rate that is close enough to the optimal sum-rate. In fact, the BB solution is certified to be at most 0.01-away from the global optimal value. The sum-rates for all other iterative methods are calculated after the sum-rates converge to their respective steady-state levels. The iterative procedure stops as the increase  in objective value between two successive iterations is $\le \epsilon$  (=0.01). We see that the suboptimal solutions achieved by our algorithm and other iterative techniques such as SPCA-SOCP, WMMSE are indeed very close to the optimal sum-rate. However, the BB method and WMMSE algorithm have slower convergence.
% The per iteration computational complexity of our proposed algorithm for a single cell $m$ is $O\left( \sum\limits_{k=1}^{K}{ D({{\mathcal{N}}_{km}}) {{N}_{\rm{t}}}} \right)$, where $D({{\mathcal{N}}_{km}})$ defines ${{\mathcal{N}}_{km}}$'s cardinality. Moreover, the KKT matrix of our proposed convex approximation is sparse, i.e., the KKT matrix consists of a lot of zeros and the sparsity of the KKT matrix increases through the iterative process, and hence, accelerates the convergence. For WMMSE, the computational complexity per iteration depends on the problem. If we consider per-antenna transmit power constraints per cell, the complexity of WMMSE is equal to that of our algorithm. However, AM and WMMSE require a large number of iterations to reach their respective suboptimal levels. For optimal BB method, with 0.01, the gap tolerance between lower and upper bound, it takes almost 900 iterations. 
%\vspace{-4mm}
  %----------------------------------------------------------------------------------------------- 
 \begin{figure}
  \centering
   \includegraphics[scale=0.18]{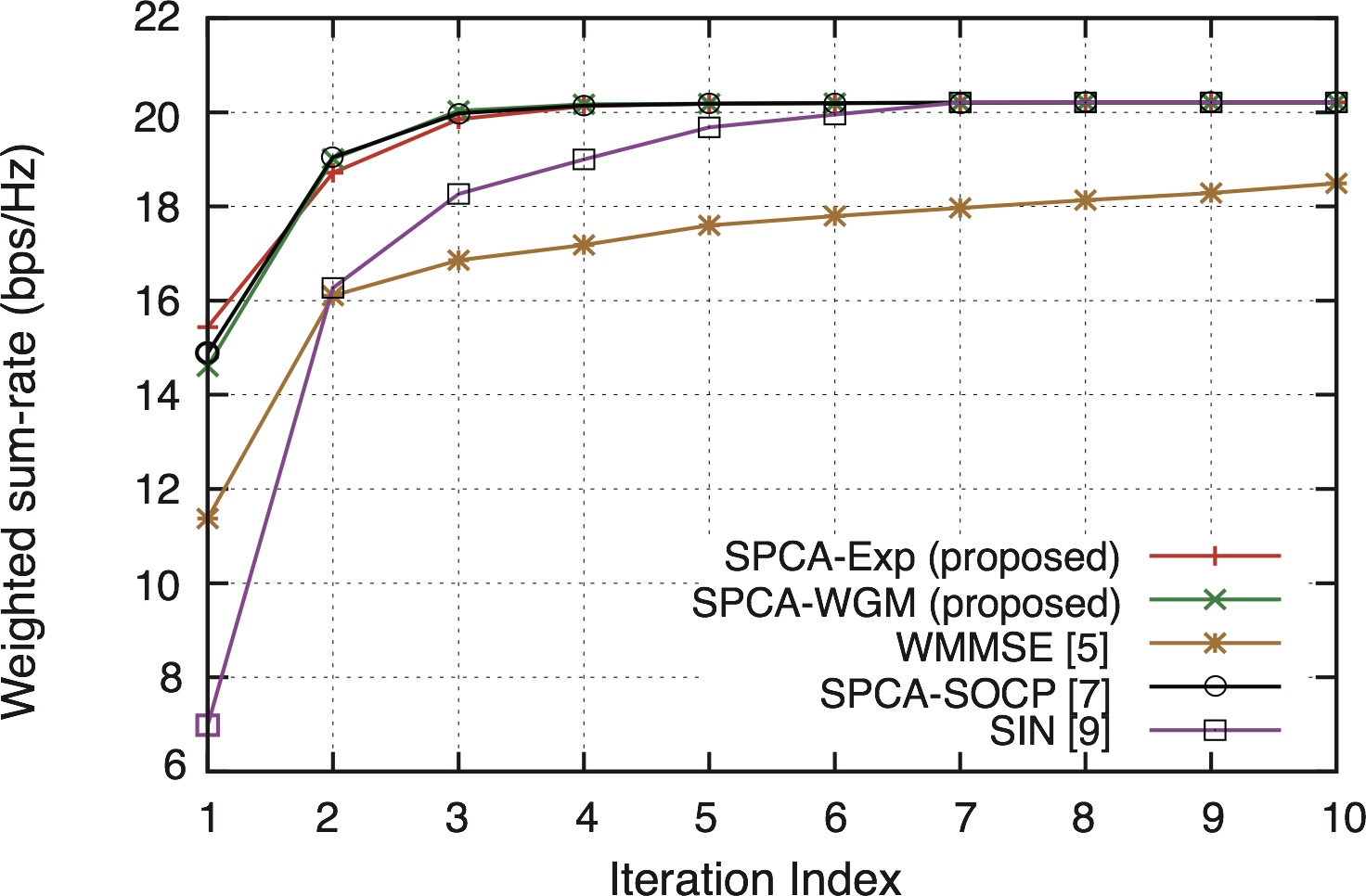}
   \caption{ Convergence behaviors of different WSRMax algorithms. }
   \label{BER2}
 \end{figure}
 %-----------------------------------------------------------------------------------------------

We compare the convergence performances of several iterative local optimal WSRMax techniques in Fig.~\ref{BER2}. The user weight $\alpha_{k,m}$ is considered as $\alpha_{k,m}\in\{0.10\sim 0.60\}, \forall k,m$. Weighted sum-rates for different methods are plotted as a function of the number of iterations required to attain respective steady outputs for a random channel realization. We see that both of our proposed methods converge to the local optimal solution within few iterations. The faster convergence of our proposed WSRMax algorithm may be attributed to the fact that the KKT matrix in each iteration is a sparse matrix, i.e., a lot of zeros appear in the KKT matrix. Operations using sparse-matrix structures and algorithms are relatively fast. The SIN method in \cite{Chris} also shows good convergence performance. For some channel realizations, the convergence behaviors of SIN and our methods are very similar. The sum-rate for the iterative method is calculated after the solution converges. 
%----------------------------------------------------------------------------------------------- 
 \begin{figure}
  \centering
   \includegraphics[scale=0.18]{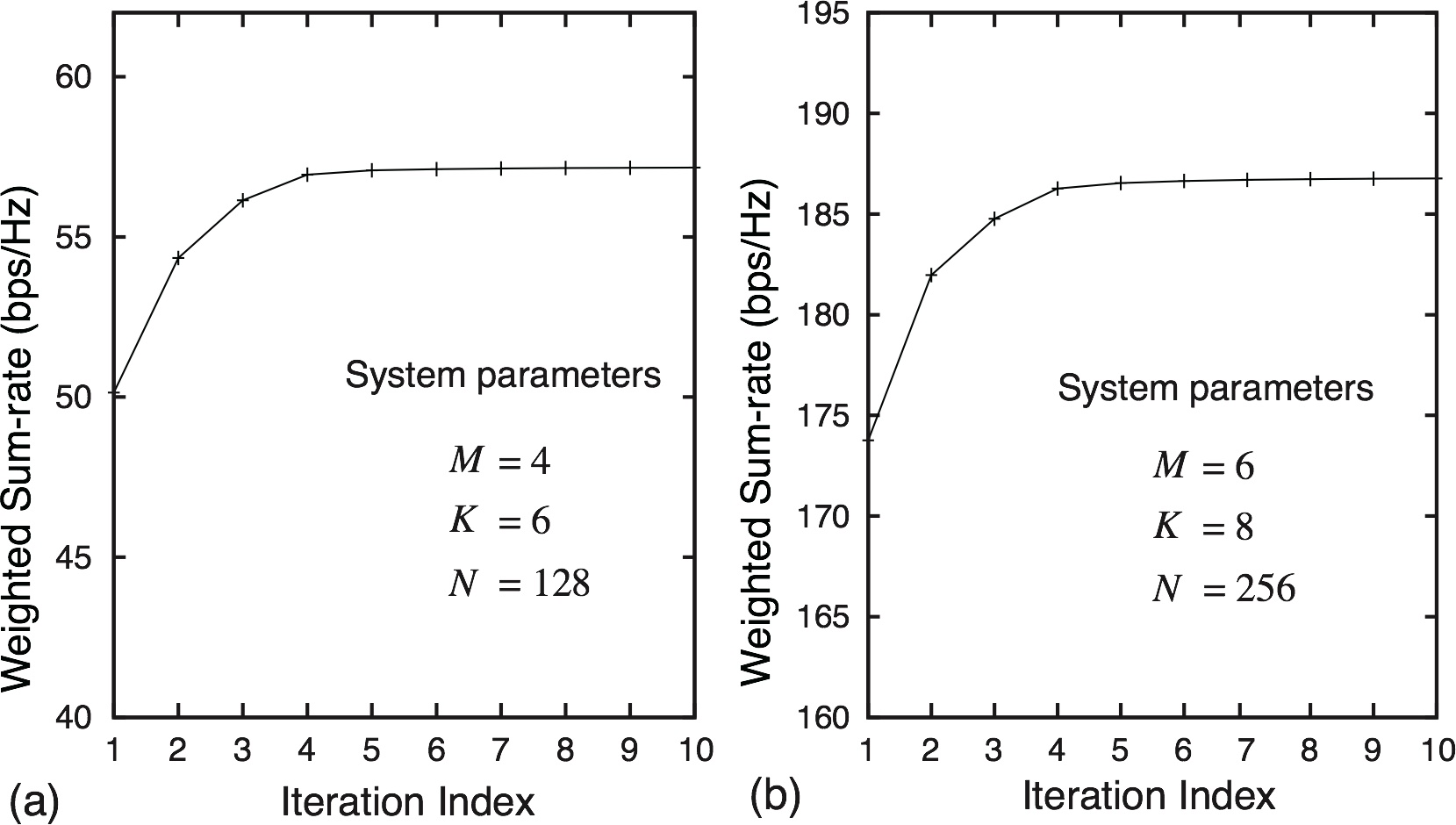}
   \caption{ Convergence behavior of the proposed solution SPCA-Exp with larger systems. }
   \label{BER3}
 \end{figure}
 %-----------------------------------------------------------------------------------------------
%The SPCA-SOCP scheme has also a very good convergence rate. However, the computational efficiency of our proposed method is marginally higher. This is due to the fact that SPCA-SOCP scheme comes with a large set of constraints along with a differential operator for linear approximation operation and looping operation to transform a function into concave form. The WMMSE and AM methods require a large number of iterations to deliver stabilized output due to their operational mechanism we discussed in Section. 1. However,  good initial values for the variables involved in WMMSE accelerate the convergence rate. 
%Though SIN algorithm, which is also based on convex approximation of the precoder covariance matrices, has similar convergence performance to our proposed method, the per iteration running time is about 7-8 times higher.

 To evaluate the convergence performance of our proposed WSRMax algorithm in more practical scenarios, we consider higher values of the system parameters and depict the simulation results in Fig.~\ref{BER3}. Two different cases are considered as shown in Fig.~\ref{BER3}a and Fig.~\ref{BER3}b, respectively. We notice that the convergence behaviors almost remain the same in terms of the number of iterations. However, the per-iteration running time increases linearly as the values of the simulation parameters such as $M$, $N$ and $K$ increases. Therefore, we can claim that the proposed WSRMax solutions are linear-time method, i.e., the complexity per cell per-iteration grows linearly with the values of $M$, $N$ and $K$.
%----------------------------------------------------------------------------------------------- 
 \begin{figure}
  \centering
   \includegraphics[scale=0.18]{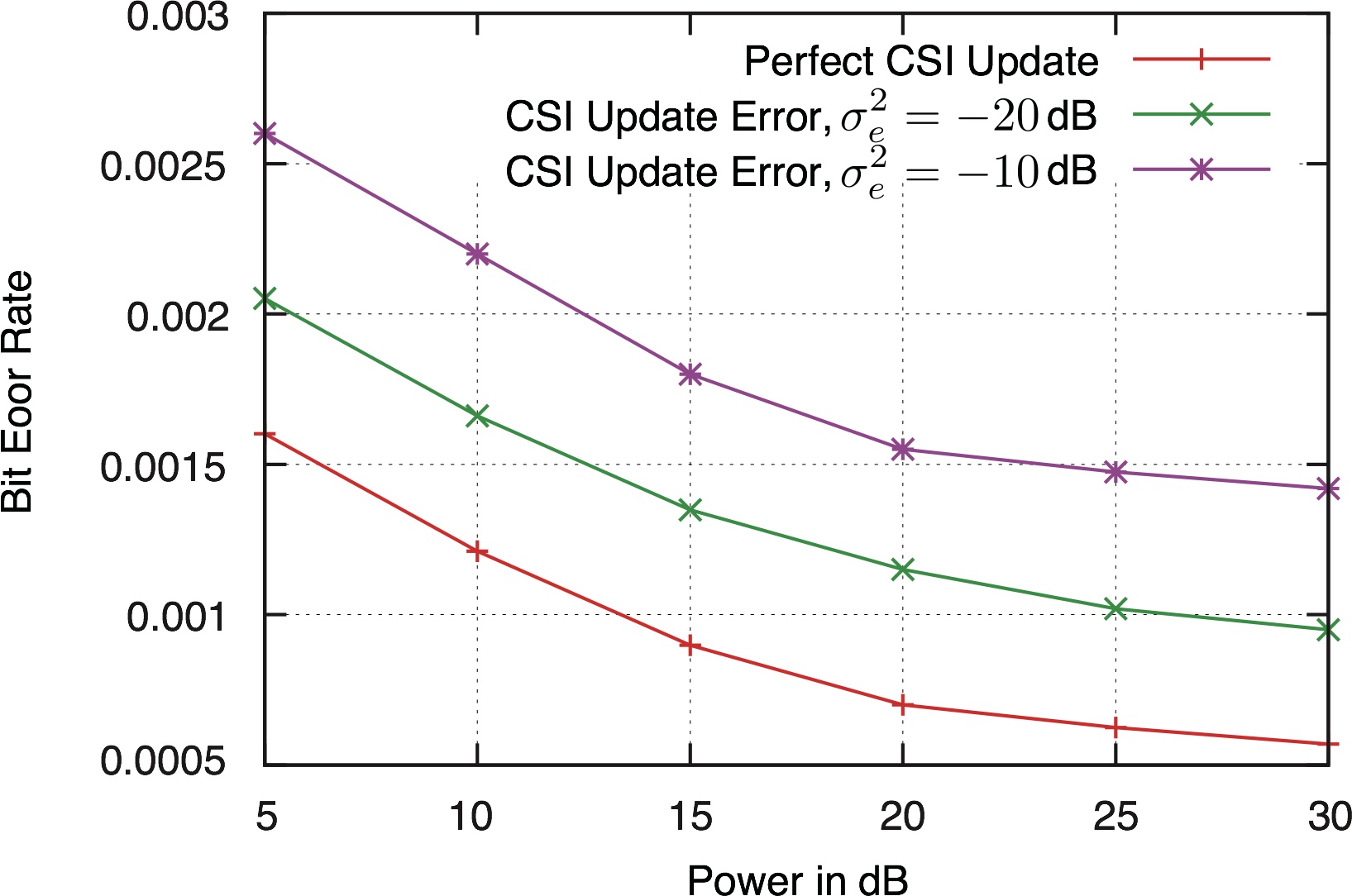}
   \caption{ BER performance of the proposed solution SPCA-Exp with CSI update error for a system with $M=2$, $K=2$, and $N=64$. }
   \label{BER4}
 \end{figure}
 %-----------------------------------------------------------------------------------------------

 To investigate the impact of CSI update error, we plot the bit error rate (BER) performance curves for mismatched beamforming scenarios due to CSI update error with varying variances in Fig.~\ref{BER4}. The CSI update error for the channel coefficient vector of user $k$ over subcarrier $n$, ${\bm{h}}_{k,m,n}$ is defined as ${\tilde{\bm{h}}}_{k,m,n}={\bm{h}}_{k,m,n}+{\bm{e}}_{k,m,n}$, where the elements of error vector ${\bm{e}}_{k,m,n}$ are distributed as $\mathcal{CN}(0,{\sigma}_e^2)$. We notice that the BER performance degrades as the variance of error is increased. In a matched beamforming situation (perfect CSI update), for any particular user, the optimizer finds the optimal beamformer and power allocation that maximizes the  system throughput by minimizing the user's leakage interference to other users. As a results, the users are able to detect the transmitted symbols intended for them correctly. Whereas in mismatched beamforming (due to imperfect CSI update) scenarios, because of higher leakage interference from other users operating on the same subcarrier, the BER performance degrades.

In temporally correlated fading scenarios under the effect of feedback delay, the CSI as well as the beamformers update frequency depends on the Doppler frequency, transmit frame duration and the CSI error tolerance. The wide-sense stationary (WSS) Jakes' statistical model defines the temporal autocorrelation function as $\rho=\mathbb{E}\{\bm{h}_{k,m,n}^{\rm{H}}(t){{\bm{h}}_{k,m,n}}(t+1)\}={{J}_{0}}(2\pi f_{\rm{D}}T_{\rm{f}}),{{\forall }{k,m}}$,
%\begin{equation}
%\label{corr}
%\rho=\mathbb{E}\{\bm{h}_{k,m}^{\rm{H}}(t){{\bm{h}}_{k,m}}(t+1)\}={{J}_{0}}(2\pi f_{\rm{D}}T_{\rm{s}}),{{\forall }{k,}}
%\end{equation}
where $t$ is the frame index, $f_{\rm{D}}$ is the Doppler frequency, $T_{\rm{f}}$ is the frame time duration, and ${J}_{0}$ is the Bessel function of the first kind of zero-th order. The quantity, $f_{\rm{D}}T_{\rm{f}}$ is generally defined as normalized Doppler frequency. We have found that for a system with CSI error tolerance, ${\sigma}_e^2=-20$ dB and $f_{\rm{D}}T_{\rm{f}}=0.01$ ($T_{\rm{f}}$=10\hspace{.5mm}msec and $f_{\rm{D}}$=1\hspace{.5mm}Hz), the CSI needs to be updated within the time span equivalent to several frames duration. However, in highly correlated fading scenarios, i.e., $f_{\rm{D}}T_{\rm{f}}=0.001$ ($f_{\rm{D}}$=0.1\hspace{.5mm}Hz) with the same error tolerance, a less frequent CSI update becomes realizable. If we set the CSI error tolerance to -10 dB, then for a system with with $f_{\rm{D}}T_{\rm{f}}=0.01$, the CSI can be updated less frequently when compared to the case with ${\sigma}_e^2=-20$ dB. Therefore, the frequency of CSI update can be several frames depending on the system parameters.

From the point of view of computational complexity, our proposed algorithm is computationally more efficient compared to the other WSRMax methods discussed, such as WMMSE, SIN. The SIN method is based on solving determinant-maximization (MAXDET) programs. So the computational complexity is high when the BS has large number of antennas. The per-iteration running time of SIN is 5-6 times higher compared to our proposed algorithm. The WMMSE admits an water-filling solution if a sum-power constraint per-BS is considered, and a closed form solution can be found per iteration. If we consider a per-antenna transmit power constraint at each BS, the complexity in per-iteration is same as that of our algorithm. However, the convergence speed of WMMSE is slow since it is based on alternating optimization concept. Therefore, we can claim that our proposed WSRMax algorithm requires lower complexity. The computational complexity of SPCA-SOCP is same as our proposed solution SPCA-WGM, also the per-iteration running times are equal because the CVX transforms the WGM function into the equivalent SOCP form before solving the problem.

%\begin{table}[h]
% \centering
%\caption{Convergence rate comparision} 
%\label{CDIopt}
%\begin{tabular}{|@{}c@{}|c@{}|c@{}|c@{}|c@{}|c@{}|c@{}|}
%\hline
%\multicolumn{2}{|c|}{Optimization Method}  & AM & SIN & SPCA-SOCP & BB & Proposed  \\
%\cline{1-7}
%\multicolumn{2}{|c|}{Per-Iteration Running Time}  & 0.312 & 4.13 & 0.97 & 0.77 & 0.94  \\
%\cline{1-7}
%\multicolumn{2}{|c|}{Approx. Iterations Required}  & 170 & 9 & 8 & 900 & 8  \\
%\cline{1-7}
%\end{tabular}
%\end{table}
%
Numerical results reveal that our proposed approach SPCA-Exp seems to have lower modeling time taken by the convex optimization tool compared to that of the approach in \cite{Mirza}. This may be due to the fact that the algorithm in \cite{Mirza} employs a differential operator and linear approximation technique. Furthermore, the approach relies on a looping operation for scaling up the user weights so that a specific set of constraints become concave although the scaling can be done easily by a single multiplication, i.e., multiplying all weighting factors with $2/{\rm{min}}(\alpha_{k,m}, k={1,2,\cdots,K},\forall m)$. Therefore, the convex approximation in \cite{Mirza} comes up with more complex problem formulation, and compared to \cite{Mirza}, the proposed approaches are more efficient. The optimal BB method is computationally very expensive. It approximately takes 900 iterations with a gap tolerance of 0.01. If we run the BB method with a gap of 0.001 or even smaller, it takes a huge number of iterations and much longer time to converge.

\section{Conclusions}
\label{section:chapter4-5}
In this letter, we study the WSRMax optimization problem for a MC-MU-OFDM downlink beamforming system. We formulate and propose two fast, provably convergent, computationally efficient SPCA based convex approximation algorithms with less-complex problem formulation. These iterative optimization solutions are convergent to the local optimal solutions. The algorithms exhibit excellent performance and outperform some previously analyzed iterative solutions for the WSRMax problem, in terms of convergence rate and computational efficiency. The most significant benefit of our proposed approaches is that they are general enough to apply to a variety of problems relating to SINR optimization. The faster convergence behavior of our proposed WSRMax algorithms may also be useful for distributed implementation.

%\ifCLASSOPTIONcaptionsoff
%  \newpage
%\fi

%===========================================

\end{document}